\documentstyle[aps,twocolumn,epsfig]{revtex}\begin{document}

\twocolumn[\hsize\textwidth\columnwidth\hsize\csname@twocolumnfalse\endcsname

\title{Fermionic entanglement in itinerant systems}
\author{Paolo Zanardi and Xiaoguang Wang}
\address{Institute for Scientific Interchange (ISI) Foundation, 
Viale Settimio Severo 65,I-10133 Torino, Italy}
\date{\today}
\maketitle

\begin{abstract}
We study pairwise quantum entanglement in  systems of  fermions itinerant
in a lattice  from a second-quantized perspective.
Entanglement in the grand-canonical ensemble is studied, both for energy eigenstates
and for the thermal state.
Relations between entanglement and superconducting correlations are discussed
in a   BCS-like model and for $\eta$-pair superconductivity.
\end{abstract}

\draft

\pacs{PACS numbers: 03.67.Lx, 03.65.Fd}
] 

\section{introduction}

The concept of quantum entanglement \cite{Schr} is believed to play an
essential role in quantum information processing (QIP) \cite{QIP}. As a
consequence many efforts have been devoted to the characterization of
entanglement \cite{Entang}. The very definition of entanglement relies on
the tensor product structure of the state-space of a composite quantum
system. Unfortunately, due to quantum statistics, such a structure does not
appear in an obvious fashion for systems of indistinguishable particles,
i.e., bosons or fermions. Indeed for these systems, in view of the (anti)symmetrization
postulate,
 one has to restrict to a subspace of the $N$-fold tensor product
of the single particle spaces. Such a subspace, e.g., the totally
anti-symmetric one, has not a naturally selected tensor product structure.
It turns out that the notion of entanglement is affected for systems of
indistinguishable particles by some ambiguity.

Since it is of direct relevance to several implementation
proposals for QIP e.g., quantum-dots based,
this issue has been very recently addressed in the literature \cite
{Schli,Li,Paolo,You,Fish}. A quantum computation model was proposed\cite{Bravyi}
by using $L$ local fermionic modes (LFMs)-- sites which can be either empty
or occupied by a fermion. Moreover the use of quantum statistics for some
QIP protocols have been analyzed \cite{bose}.

Along the same line of realizing a bridge between quantum information science
and conventional many-body theory it has been  discussed
 entanglement in magnetic systems \cite
{Connor01,Meyer01,Arnesen01,Wang01,WangKlaus}.
In particular entanglement in both the ground state \cite
{Connor01,Meyer01} and thermal state \cite{Arnesen01,Wang01,WangKlaus} of a
spin-1/2 Heisenberg spin chain have been analyzed in the literature. 
In this situation the system state is given by the Gibb's density operator $%
\rho _T=\exp \left( -H/kT\right) /Z,$ where $Z=$tr$\left[ \exp \left(
-H/kT\right) \right] $ is the partition function, $H$ the system
Hamiltonian, $k$ is Boltzmann's constant which we henceforth will take equal
to 1, and $T$ the temperature. As $\rho (T)$ represents a thermal state, the
entanglement in the state is called {\em thermal entanglement}\cite
{Arnesen01}.
Finally the  intriguing issue of the relation between entanglement and
quantum phase transition \cite{QPT} have been  addressed  in a few quite recent
papers \cite{Osborne,Osterloh}. 

In this paper we will explore the relations between 
entanglement and (super)conducting correlations
by following the spirit of Ref. \cite{Paolo}. 
It is important to stress that, due to the lack of measure of genuine many-body
entanglement, we restrict us pairwise entanglement
in this paper.
Notice that  in this approach 
{\em the subsystems are given by modes and not by particles}. 
This is therefore an essentially {\em second-quantized} 
approach \cite{WULI}.

In Sec. II  basic definitions are given  and the mapping scheme 
between LFMs and qubits introduced in 
\cite{Paolo} is briefly recalleed. 
In Sec. III the entanglement in both  eigenstates and   thermal 
state is  studied for free fermions
hopping in a lattice. In Sec. IV the relations between pairwise entanglement of 
 and  superconducting correlations are discussed for 
Two types of superconductivity,  BCS-like superconductivity \cite{BCS},
and the so-called $\eta$-pair superconductivity\cite{Yang}.
Sec. V contains the conclusions.

\section{Lattice Fermions}

%%%%%%%%%%%%%%%%%%%%%%%%%%%%%%%%%%%%%%%%%%%%
Let us start by recalling some basic facts about (spinless) fermions on a
lattice. In the second-quantized picture the basic objects are the
creation and annihilation operators $c_l^{\dagger }$ and $c_l$ of $l$-th
LFM. They satisfy the canonical anti-commutation relations
\begin{equation}
[c_i,c_j]_{+}=0,\quad[c_i,c_j^{\dagger }]_{+}=\delta _{ij}.  \label{car}
\end{equation}
The Hilbert space naturally associated to the $L$ LFMs, known as Fock space $%
{\cal H}_F$, is spanned by $2^L$ basis vectors $|n_1,...,n_{L}\rangle:=%
\prod_{l=1}^{L} (c_l^\dagger)^{n_l}\vert 0\rangle\,(n_l=0,1\forall\, l).$

{}From the above occupation-number basis it should be evident that ${\cal H}%
_F$ is isomorphic to the $L$--qubit space. This is easily seen by defining
the mapping \cite{Paolo} 
\begin{equation}
\Lambda :=\prod_{l=1}^L(c_l^{\dagger })^{n_l}|0\rangle \mapsto \otimes
_{l=1}^L|n_l\rangle =\otimes _{l=1}^L(\sigma _l^{+})^{n_l}|0\rangle ,
\label{mappp}
\end{equation}
where $\sigma _l^{+}$ is the raising operator of $l$-th qubit.
This is an Hilbert-space isomorphism between ${\cal H}_F$ and ${\rm \kern%
.24em\vrule width.04emheight1.46exdepth-.07ex\kern-.30emC}^{\otimes \,L}$.
By means of this identification one can discuss entanglement of fermions by
studying the entanglement of qubits. Clearly this entanglement is strongly
relative to the mapping (\ref{mappp}) and it is by no means unique. By
defining new fermionic modes by automorphisms of the algebra defined by Eq.(\ref{car}) one gives rise to different mappings between ${\cal H}_F$ and $%
{\rm \kern.24em\vrule width.04emheight1.46exdepth-.07ex\kern-.30emC}%
^{\otimes \,L}$ with an associated different entanglement. This simple fact
is one of the manifestations of the relativity of entanglement \cite{Paolo2}.

It is useful to see how the mapping (\ref{mappp}) looks on the operator
algebra level. {}From the relation 
$c_l^\dagger\vert
n_1,...,n_{L}\rangle=\delta_{n_l,0}(-1)^{\sum_{k=1}^{l-1}n_k} \vert
n_1,...,n_{l-1},1,n_{l+1},...,n_{L}\rangle, $ 
it follows that 
\begin{equation}
c_l^\dagger\mapsto \sigma_l^+\prod_{k=1}^{l-1}(-\sigma_k^z),  \label{eq:jw}
\end{equation}
where $\sigma_k^z$ is the $z$ component of the usual Pauli matrices for $k$-th qubit. This algebra isomorphism is quite well-known in the condensed matter
literature and it referred to as the Jordan-Wigner transformation\cite{JW}.
Notice that the inverse of Eq.(\ref{eq:jw}) is given by $\sigma_l^\dagger%
\mapsto c_l^\dagger\prod_{k=1}^{l-1}$$\exp(i\pi c_k^\dagger c_k)$. We see
that, due to the {\em non-local} character of the mapping $\Lambda$($%
\Lambda^{-1}$) even simple fermionic (spin) models can be transformed into
non-trivial spin (fermionic) models. On the other hand the fermionic state
like $\prod_kc_k^\dagger \vert 0\rangle$ are clearly mapped by $\Lambda$
onto product qubit states. 
%{\em Quantum entanglement is then left invariant by the 
%Jordan--Wigner transformation.}

It is important to keep in mind that, for charged and/or massive fermions,
the Fock space is {\em not} the state-space of any physical system. Indeed,
at variance with massless neutral particles, e.g., photons, only eigenstates
of $\hat{N}= \sum_{l=1}^L c_l^\dagger c_l$ are allowed {\em physical}
vectors and, for the same reason, only operators commuting with $\hat{N}$
could be physical observables. This of course is nothing but a
{\em superselection rule}, i.e., ${\cal H}_F=\oplus_{N=0}^L {\cal H}_F(N)$ that
does not allow for linear superposition of states corresponding to different
charge/mass eigenvalues \cite{selection}.

Despite the above considerations we notice that in some situations one is
led to attribute to the whole Fock space some physical meaning. This happens
for systems in a symmetry broken phase. For example in superconductivity and
superfuidity the order parameter corresponds to an expectation value of an
operator connecting different $N$-sectors. It follows that the associated
mean-field Hamiltonian does not commute with $N$.

Of course one can argue that this kind of violation occurs on a level that has
not any deep physical significance, after all the mean-field approach is
just a variational one aimed to produce good approximation to physical
expectation values. According to this view therefore the properties, e.g.,
entanglement, of the ansatz states should not to be taken too seriously.
Nevertheless we think that this issue has some interest and the relations
between pairwise entanglement and superconductivity will be provided before the section of conclusions.

\section{Itinerant systems}

Let us now consider free spinless fermions in a lattice. The Hamiltonian is
given by 
\begin{equation}
H=-t\sum_{l=1}^L\left( c_l^{\dagger }c_{l+1}+c_{l+1}^{\dagger }c_l\right)
-\mu \sum_{l=1}^Lc_l^{\dagger }c_l  \label{eq:hh1}
\end{equation}
with the periodic boundary condition. Here $t$ represents the hopping
integral between sites and $\mu $ is the chemical potential.

It is known that the eigenvalue problem of $H$ can be
solved by a discrete Fourier transformation (DFT) 
\begin{equation}
c_l=\frac 1{\sqrt{L}}\sum_{k=1}^L\omega ^{lk}\tilde{c}_k,  \label{eq:ft0}
\end{equation}
where $\omega =\exp (i2\pi /L).\,$After the DFT, the Hamiltonian (\ref
{eq:hh1}) becomes
\begin{equation}
H=-2t\sum_{k=1}^L\cos (2\pi k/L)\tilde{c}_k^{\dagger }\tilde{c}_k-\mu \hat{N}%
,  \label{eq:h2}
\end{equation}
where $\hat{N}=\sum_{k=1}^L\tilde{c}_k^{\dagger }\tilde{c}%
_k=\sum_{l=1}^Lc_l^{\dagger }c_l$ is the total fermion number operator.
{}From Eq.(\ref{eq:h2}), we immediately obtain the eigenvectors

\begin{eqnarray}
|{\bf k}_N\rangle &=&\tilde{c}_{k_1}^{\dagger }\tilde{c}_{k_2}^{\dagger },...,%
\tilde{c}_{k_N}^{\dagger }|0\rangle ,\nonumber\\
{\bf k}_N&=&{\bf (}k_1,k_2,...,k_N{\bf %
)\in Z}^N,  \label{eq:eigen1}
\end{eqnarray}
and the corresponding eigenvalues
\begin{eqnarray}
E_{{\bf k}_N}=\sum_{l=1}^N(\epsilon_{k_l}-\mu), \, \epsilon_{k_l}&=&-2t\cos
(2\pi k_l/L).  \label{eq:eigen2}
\end{eqnarray}

Associated to the new fermionic modes $\tilde{c}_{k}$ there is a tensor
product structure for the Fock space. The latter is defined by the mapping 
\begin{equation}
\Lambda_{{\rm DFT}}: =\prod_{k=1}^{L}(\tilde{c}_k^\dagger)^{n_k} \vert
0\rangle\mapsto \otimes_{l=k}^{L}\vert n_k\rangle,
\end{equation}
Obviously since the eigenstates $|{\bf k}_N\rangle $ are products with
respect the tensor product structure due to $\Lambda_{{\rm DFT}}$ the entanglement in the 
eigenstates (\ref{eq:eigen1}) is always zero. However entanglement,
associated with map $\Lambda,$ may exist in the eigenstates. For instance,
the concurrence $C=2/L$ for any pair of fermions when $N=1$.
The corresponding eigenstates are called W states \cite{Koashi,Dur,Wang01}.
 
\subsection{Entanglement in the eigenstates}

%%%%%%%%%%%%%%%%%%%%%%%%%%%%%%%%%%%%%%%%%%%%%%%%%%%%%%%%%%% 
In order to make an analysis of the entanglement of our spinless fermions we will use the notion of concurrence \cite{Con}, This is a simple measure for two qubits that allows to quantify the entanglement between {\em any} pair of fermions by our mapping.

We define the reduced density matrix associated to the first and second
LFMs as $\rho^{(12)}\in $End({\bf C}$^4$). Note that the Hamiltonian
is translation invariant, therefore entanglement between nearest-neighbor
fermions are identical. Due to the fact that $[\hat{N},H]=0, $ the reduced
density matrix have the following form

\begin{equation}
\rho ^{(12)}=\left( 
\begin{array}{llll}
u &  &  &  \\ 
& w_1 & z &  \\ 
& z^{*} & w_2 &  \\ 
&  &  & v
\end{array}
\right)  \label{eq:reduce}
\end{equation}

The nonvanishing relevant matrix elements of $\rho ^{(12)}$ are given by ($
\langle \bullet \rangle $ denotes the expectation value over $\rho $), 
\begin{eqnarray}
u &=&1-2\langle \hat{N}\rangle /L+\langle \hat{n}_1\hat{n}_2\rangle , 
\nonumber \\
v &=&\langle \hat{n}_1\hat{n}_2\rangle ,  \nonumber  \\
z &=&\langle c_1^{\dagger }c_2\rangle .  \label{eq:eleee}
\end{eqnarray}
The concurrence of $\rho ^{(2)}$ is then given by \cite{Connor01} 
\begin{equation}
C=2\max \{0,|z|-\sqrt{uv}\}.  \label{eq:conccc}
\end{equation}
As the matrix elements $w_1$ and $w_2$ do not appear in the concurrence (\ref
{eq:conccc}), their expressions are not given in this paper. {}
From Eqs.(\ref{eq:eleee}) and (\ref{eq:conccc}) it follows that, in order
to obtain the
concurrence,
we need to compute the {\em correlation functions} 
$\langle \hat{n}_1\hat{n}_2\rangle $ and $\langle c_1^{\dagger }c_2\rangle $
and the mean fermionic number $\langle \hat{N}\rangle .$
{}
For the eigenstate $|{\bf k}_N\rangle $ ,
after direct calculations \cite{identity}, we obtain 
\begin{eqnarray}
u &=&(n-1)^2-|S_{{\bf k}_N}|^2,  \nonumber \\
v &=&n^2-|S_{{\bf k}_N}|^2,  \nonumber \\
z &=&S_{{\bf k}_N},  \label{eq:ele1}
\end{eqnarray}
where $n=N/L$ is the filling and 
\begin{equation}
S_{{\bf k}_N}=L^{-1}\sum_{l=1}^N\omega ^{k_l}=L^{-1}\sum_{l=1}^Ne^{ik_l2\pi
/L}.  \label{eq:skn}
\end{equation}

By combining Eq. (\ref{eq:conccc}) and (\ref{eq:ele1}) one gets the concurrence
between two LFMs, 
\begin{eqnarray}
C &=&2\max \{0,|S_{{\bf k}_N}|-  \nonumber \\
&&\{[(n-1)^2-|S_{{\bf k}_N}|^2][n^2-|S_{{\bf k}_N}|^2]\}^{1/2}\},
\label{eq:concon}
\end{eqnarray}
which is determined only the filling and the correlation function $\langle
c_1^{\dagger }c_2\rangle =S_{{\bf k}_N}.$ 
 This latter quantity is obviously related to the ``itinerancy'' of the state, i.e., how fermions
propagate. It follows that the concurrence (\ref{eq:concon}) contains direct
information about the conducting properties of the given quantum state.

{}From Eq.(\ref{eq:concon}) one can directly see that there exists entanglement
 if the correlation function $z$ and the filling factor $n$ satisfy 
the equation $|z|^4-2(n^2-n+1)|z|^2+(n^2-n)^2<0$. 
Then we obtain that there exists 
pairwise entanglement between two LFMs in the eigenstates if $|z|^2$ in the range   
$n^2-n+1-\sqrt{2n^2-2n+1}<|z|^2< n^2-n+1+\sqrt{2n^2-2n+1}$. 
In the case of $N=1$, $|S_{{\bf k}%
_N}|=n=1/L$, and therefore $C=2/L$\cite{Koashi,Dur,Wang01}. For $L=2,$ this state is maximally entangled. A trivial case is $N=0$, there is no entanglement at all. Now we
consider the case $N=L$, i.e, the lattice is fully filled. Now $|S_{{\bf k}%
_N}|=0$, and hence $C=0$. For the half filling ($n=1/2$), $u=v$ and Eq.(\ref{eq:concon}%
) reduces to $C=2\max \{0,|S_{{\bf k}_N}|+|S_{{\bf k}_N}|^2-1/4\}.$ Then
the entanglement exists if and only if   $(\sqrt{2}-1)/2<|S_{{\bf k}_N}|\le 1/2$.

Now we consider the ground-state $|G\rangle$ of the system. It is  obtained by
filling the lowest single particle
energy levels. By taking for simplicity $N$ odd one has
$|G\rangle=c_{k_0}^\dagger\prod_{n=1}^{N/2-1} c_{k_n}^\dagger c_{-k_n}^\dagger \,|0\rangle
(k_n=2\pi n/L).$
From this definition  it follows
$S_G=L^{-1}(1+2\Re \sum_{n=1}^{N/2-1} \omega^n)=
2\,L^{-1} [\cos(\pi(N/2-1)/L)\sin((N/2-1))/\sin(\pi/L)-1],$ now we take the limit $L\mapsto\infty,\,
N/L\mapsto n,$ the resulting expression is given by
\begin{equation}
S_G(n)=\frac{1}{\pi}\sin(\pi n)
\end{equation}
where we have used $\lim_{L\rightarrow \infty }L\sin (\pi /L)=\pi .$
Moreover from the  inequality $S_G\ge n\,(1-n)$ it follows
that the second argument of the $\max$ function in Eq. (\ref{{eq:concon}})
it is always non negative and we can get rid of the maximization.
the concurrence then becomes

\begin{eqnarray}
C &=&2\, \{\sin (\pi n)/\pi -[(n-1)^2-\sin ^2(\pi n)/\pi ^2]^{1/2}\}\nonumber\\
&&\times[n^2-\sin ^2(\pi n)/\pi ^2]^{1/2}\},\label{eq:infinity}
\end{eqnarray}
in the infinite lattice. Fig. 1 shows the concurrence as a function of the
filling $n$ in the infinite lattice. We see that the entanglement becomes
maximal at half filling for two neighboring fermions, and the entanglement
is symmetric with respect to the point of half filling. At the point of half
filling the concurrence simply becomes $C=2/\pi +2/\pi ^2-1/2\approx
0.339262.$ The property of symmetry can also be  seen directly from Eq.(\ref{eq:infinity}) as the concurrence is invariant if we make the transformation $n\mapsto 1-n$.

For zero chemical potential i.e., half-filling ,
 it exists a direct 
relation  between the concurrence  and the ground state energy density $\epsilon_0$.
Indeed from translational symmetry of the hamiltonian
one has $\langle c_1^\dagger c_2 \rangle=-1/2tL \langle H\rangle=  -1/2t \epsilon_0(n),
$ where we have used even the reality of $\langle c_1^\dagger c_2.$
This relation, that can be extend to finite temperature as well, 
is in a sense remarkable in that it connects entanglement 
with a thermodynamical quantity that depends on just the partition function of system.
The latter is  determined just by the Hamiltonian spectrum, whereas computing concurrence
requires in general also the knowledge of the eigenstate.
This kind of connection between entanglement  and thermodynamical
quantities has been discussed even for spin chains, both zero \cite{connor01}
and finite temperature \cite{xwz3}.

\subsection{ Thermal Entanglement}

In this section we extend our analysis to the entanglement in a finite temperature. The
state of fermions at thermal equilibrium is described by the following
Gibbs gran-canonical state 
\begin{equation}
\rho _T=\frac 1Z\sum_{{\bf k}_N}\exp \left( -\beta E_{{\bf k}_N}\right) |%
{\bf k}_N\rangle \langle {\bf k}_N|.  \label{eq:rhoo}
\end{equation}
where $\beta =1/kT$, $k$ is the Boltzman's constant, and the partition
function $Z$ is given by

\begin{equation}
Z=\sum_{{\bf k}_N}\exp \left( -\beta E_{{\bf k}_N}\right)
=\prod_{k=1}^L[1+e^{-\beta (\epsilon _k-\mu )}].
\end{equation}
The average accupation numbers $\langle n_k\rangle $ is given by

\begin{equation}
\langle n_k\rangle_T =\frac 1{e^{\beta (\epsilon _k-\mu )}+1}, 
\end{equation}
which is the Fermi-Dirac distribution. The expecation value $\langle \hat{N}%
\rangle $ is then easily obtained as $\langle \hat{N}\rangle
=\sum_{k=1}^L\langle n_k\rangle .$

The reduced density matrix $\rho _T^{(12)}$ associated with $\Lambda _{{\rm 
DFT}}$ will be a $4\times 4$ diagonal matrix. Then from Eq.(\ref{eq:conccc}),
the concurrence is zero for any pair of LFMs. So we will discuss the
entanglement associated to the map $\Lambda $. {}The form of the reduced
density matrix in the state $\rho _T$ is then given by Eq.(\ref{eq:reduce})
and the concurrence is given by Eq.(\ref{eq:conccc}). Now we need to
calculate the correlation functions in Eq. (\ref{eq:eleee}) for the state $\rho _T.$
The result is readily obtained from Eq. (\ref{eq:ele1})
by replacing the eigenvalues of the $n_k$'s with the corresponding thermal averages: 
\begin{eqnarray}
\langle c_1^{\dagger }c_2\rangle &=&L^{-1}\sum_{k=1}^L\omega ^k\langle
n_k\rangle .  \nonumber \\
\langle n_1 n_2\rangle &=&\langle \hat{N}\rangle ^2/L^2-|\langle c_1^{\dagger
}c_2\rangle |^2.  \label{eq:eee}
\end{eqnarray}

Then the exact expression for the concurrence is  given by the
combination of Eqs.(\ref{eq:eleee}), (\ref{eq:conccc}), and (\ref{eq:eee}). Note
the  concurrence is thus  obtained in an analytical form for arbitrary $N.$
To exemplify this result let us  first consider
 the simple case of $L=2$
 For  two sites, from Eqs.(\ref{eq:eleee}), 
(\ref{eq:conccc}), and (\ref{eq:eee}), the concurrence is given by
\begin{equation}
C=\frac{\max \{0,\sinh (2\beta |t|)-1\}}{\cosh (\beta \mu )+\cosh (2\beta t)}%
,  \label{eq:c22}
\end{equation}
which is similar to the concurrence in a thermal state of the two-qubit $XX$
model \cite{Wang01}. For three special values of $\langle \hat{N}\rangle $,
we have table I, from which we see that there is no entanglement when the chemical potential is $\mp\infty$ (the corresponding mean fermion number is 0 and 2).
The entanglement is maximal when $\mu=0$ and other parameters are fixed.
The mean fermion number is simply given by 
\begin{equation}
\langle \hat{N}\rangle =\frac{e^{\beta \mu }+\cosh (2\beta t)}{\cosh (\beta
\mu )+\cosh (2\beta t)},  \label{eq:N}
\end{equation}
from which we obtain 
$
\mu =\beta ^{-1}\ln \{(2-\langle \hat{N}\rangle
)^{-1}\{\cosh (2\beta t)(\langle \hat{N}\rangle -1)+[\cosh ^2(2\beta
t)(\langle \hat{N}\rangle -1)^2+2\langle \hat{N}\rangle -\langle \hat{N}%
\rangle ^2]^{1/2}\}\}.
$
{}From this relation and Eq. (\ref{eq:c22}) one can calculate the
concurrence as a function of temperature and the mean fermion number $%
\langle \hat{N}\rangle .$ This function is represented in Fig.2. We observe
that the entanglement becomes maximal when the mean fermion number is 1 for
fixed temperature $T$. Let us first comment on the $n\mapsto 1-n,$ or in
view of Eq.(\ref{eq:c22}) equivalently $\mu \mapsto -\mu ,$ symmetry of the
function $C$. This fact can be understood from Eq.(\ref{eq:conccc}).
Clearly the only thing to check is that $|z(t,\mu )|=|z(t,-\mu )|.$ This
latter statement easily follows by the use of particle-hole transformation
i.e., $c_j\leftrightarrow c_j^{\dagger }$ $(j=1,\ldots ,L)$ realizing $%
z(t,-\mu )\mapsto z(-t,-\mu )$ along with the unitary mapping $c_j\mapsto
(-1)^j\,c_j$ that, for bi-partite lattices, changes the sign of $t.$ Another
important feature is the existence of a threshold temperature, after which
the entanglement disappears. Remarkably this threshold temperature is
independent of the mean fermion number. This phenomenon is related to the
independence on an external magnetic field displayed by the associated spin model 
\cite{Arnesen01,Wang01}.

For large $L$, we plotted in Fig. 3 the concurrence as a function of
temperature for different $\mu$. We observe that the threshold temperature
is independent of $\mu$. Moreover it is worth noticing that, for
sufficiently high $\mu$ ,i.e, filling, one has a non monotonic behavior of
the concurrence as a function of $T$. Indeed we see that the entanglement
can initially {\em increase} as the temperature is raised. This
phenomenon is due to the fact that the chemical potential in fermionic
systems plays a role analogous to the external magnetic field $B_z$ for spin
systems \cite{Arnesen01}. When $B_z$ is large enough the ground state is given by a product state (all the spins aligned), hence entanglement in the thermal state is due to the excited eigenstates. Of course for $T$ large enough one always gets $C=0$ in t
hat the Gibbs state is approaching the maximally mixed state.

\section{Entanglement and superconductivity}
In this section we discuss fermionic entanglement in simple   
superconducting systems and explore the relations between 
 entanglement and superconducting correlations.
Let us first consider  a BCS-like i.e., mean-field, model

\subsection{BCS-like superconductivity}

The following BCS-like Hamiltonian  describes the
pairing between fermions carrying momentum $k$ and $-k$, $H=\sum_kH_k,$ 
\begin{equation}
H_k=\epsilon _k\,(n_k+n_{-k}-1)+\Delta _k\,c_k^{\dagger }c_{-k}^{\dagger }+%
\bar{\Delta}_k\,c_kc_{-k}.  \label{bcs}
\end{equation}
The quantities $\Delta _k=|\Delta _k|\,e^{i\,\phi _k}$ are order parameters
of conductor-superconductor phase transitions. They are determined by the
self-consistent relations $\Delta _k=\langle c_kc_{-k}\rangle $. and above
a critical temperature they vanish thus signaling the absence of
superconducting correlation.

The structure of Eq. (\ref{bcs}) clearly suggest that the relevant
tensor-product structure of this problem is given by ${\cal H}_F\cong
\otimes_k (h_k\otimes h_{-k}),$ where $h_k:=\mbox{span} \{
(c_k^\dagger)^\alpha\,|0\rangle\,/\,\alpha=0,1\}.$ Moreover it is very
simple to check that, for any $k,$ the operators $J_k:=
n_k+n_{-k}-1,\,J^+_k:=c_k^\dagger c_{-k}^\dagger,\, J_k^-:=c_{-k}c_k$
span a $su(2)$ Lie-algebra. The states $|\alpha\alpha\rangle:=
|\alpha\rangle_k\otimes |\alpha\rangle_{-k}\,(\alpha=0,1),$ realize a spin-$%
1/2$ representation of such an algebra. Therefore the Hamiltonian (\ref{bcs}%
), that it is equivalent to a spin-1/2 particle in an external magnetic
field along the $x$ direction, is readily diagonalized. For example, if we
define $\theta_k:= \tan^{-1} |\Delta_k|/\epsilon_k,$ the ground state is
given by $\otimes_k |-\rangle_k$ where 
\begin{equation}
|-\rangle_k:= \cos (\theta_k/2) \,|00\rangle_k-
e^{i\phi_k}\,\sin(\theta_k/2)\,|11\rangle_k.  \label{bcs-gs}
\end{equation}
It corresponds to the eigenvalue $E_{k-}=-E_k$. Here $E_k=\sqrt{\epsilon_k^2
+|\Delta_k|^2}.$ The self-consistent equation for the $\Delta_k$'s reads 
\begin{equation}
|\Delta_k|=\frac{\sinh(\beta\,E_k)\sin(\theta_k)}{2[\cosh(\beta E_k)+1]}
\end{equation}
The concurrence associated to the thermal state $\rho_k(\beta):=\exp(
-\beta\,H_k)/Z$ is given by 
\begin{equation}
{C} =\frac{\mbox{max}\{0,\,\sinh(\beta\,E_k)\sin(\theta_k)-1\} }{%
\cosh(\beta\,E_k) +1}  \label{bcs-con}
\end{equation}
In the limit of $T\rightarrow 0$, the concurrence becomes $\sin(\theta_k)$
which is just the concurrence of the ground state $|-\rangle_k$. 
By solving
these equations the numerical results are given in Fig.4. One can see that the concurrence goes to zero at a
temperature slightly {\em lower} than superconductor critical one. Since in
the temperature range $[0,\,T_c]$ the function $\Delta_k(T)$ is invertible
one can express the concurrence as a function of the order parameter only.
This is illustrated in Fig. 5 which shows that it is necessary to have a certain
amount of superconducting correlation in order to have an pairwise entangled thermal state. Notice that we set $\epsilon_k=0$ that in a grand-canonical picture ($\epsilon_k\mapsto\epsilon_k-\mu)$ corresponds to half-filling \cite{hf}.

We would like to observe now that a 
{\em non } mean-field Hamiltonian formally analogous to (\ref{bcs}) plays an
important role in the excitonic proposal for QIP by Biolatti {\it et al} 
\cite{bio}. In that case the fermionic bilinear terms $c_k^{\dagger
}\,c_{-k}^{\dagger }$ are replaced by $X_i:=c_l^{\dagger }\,d_m^{\dagger
}\,(i:=(l,m))$ where $c_l^{\dagger }$ ($d_m^{\dagger }$) creates an electron
(hole) in the $l$-th ($m$-th) state of the conduction (valence) band of a
semiconductor. The order parameter $\Delta _k$ becomes a (independently
controllable) coupling to an external laser field. The excitonic index $l$
can be associated to $L$ different, spatially separated, quantum dots; this
implies that $l\neq l^{\prime }\Rightarrow [X_l,\,X_{l^{\prime }}]=0.$ Let $%
|0\rangle $ denotes the particle-hole vacuum (ground state of the
semiconductor crystal). Since $X_l^2=0$ one immediately sees that the
``excitonic Fock space'' span$\{\prod_lX^{n_l}\,|0\rangle \,/\,n_l=0,1\}$ is
isomorphic to a $L$ qubit space \cite{WULI}. This example shows that one can
consider spaces allowing for a varying number of ``particles'' that
nevertheless are fully legitimate quantum state-spaces. No super-selection
rule violation (possibly due to spontaneous symmetry-breaking) has to be
invoked. Next we consider another kind of , {\em non} mean-field superconductivity, 
i.e., the $\eta$-pair superconductivity.

\subsection{$\eta $-pair superconductivity}

Yang\cite{Yang} discovered a class of eigenstates of the Hubbard model which
have the property of off-diagonal long-range order (ODLRO)\cite{Order},
which in turn implies the Meissner effect and flux quantization \cite
{Meissner}.
Let us  begin  by introducing the $\eta $-operators

\begin{equation}
\eta =\sum_{j=1}^Lc_{j\uparrow }c_{j\downarrow },\text{ }\eta
^{+}=\sum_{j=1}^Lc_{j\downarrow }^{\dagger }c_{j\uparrow }^{\dagger },\text{ 
}\eta ^z=\frac L2-\sum_{j=1}^Ln_j, 
\end{equation}
which form the su(2) algebra and satisfies $[\eta ,$ $\eta ^{\dagger
}]=2\eta ^z,[\eta ^{\pm },\eta ^z]=\pm \eta ^{\pm }.$ Here the fermions have spins and the 
operator $n_j=c_{j\downarrow}^\dagger c_{j\downarrow}+c_{j\uparrow}^\dagger c_{j\uparrow}$.
The operators $\eta^{\pm }$ also satisfy the relations $(\eta ^{\pm })^{L+1}=0,$ which
reflect the Pauli principle, i.e., the impossibility of occupying a given
site by more than one pair $c_{j\downarrow }^{\dagger }c_{j\uparrow
}^{\dagger }.$ 

In this context the relevant state-space is the one spanned by the 
$2^L$ basis vectors
\begin{equation}
|n_1,...,n_L\rangle =\prod_{l=1}^L(c_{l\downarrow }^{\dagger }c_{l\uparrow
}^{\dagger })^{n_l}|\text{vac}\rangle (n_l=0,1\text{ }\forall l).
\end{equation}
{}From the above basis it is evident that the space span$\{|n_1,...,n_L\rangle
\}$ is isomorphic to the $L$--qubit space. This is easily seen by defining
the mapping

\begin{equation}
\Lambda' :=|n_1,...,n_L\rangle \mapsto \otimes _{l=1}^L|n_l\rangle =\otimes
_{l=1}^L(\sigma _l^{+})^{n_l}|0\rangle .  \label{mapp}
\end{equation}

Then we can produce $L$ `number' state $|N\rangle $ by
applying successive powers of $\eta ^{+}$ on the vacuum state defined by $%
\eta |0\rangle =0.$ So

\begin{equation}
|N\rangle ={\cal N}^{-1/2}\eta ^{+N}|0\rangle ,%
\text{ }N=0,...,L,  \label{eq:number}
\end{equation}
where ${\cal N}= L!N!/(L-N)!.$
The span of the $|N\rangle$'s ($N=0,\ldots,L), $ known as $\eta$-paired states,
forms an irreducible spin-$L/2$ representation space. A mentioned above what makes the
number states interesting is the fact that they  have been shown to have ODLRO. 
{}Indeed from Eq.(\ref{eq:number}) the following, distance independent, 
correlation function is  obtained 

\begin{equation}
{\cal O}_{N,L}=\langle N|c_{j\downarrow }^{\dagger }c_{j\uparrow }^{\dagger
}c_{l\uparrow }c_{l\downarrow }|N\rangle =\frac{N(L-N)}{L(L-1)}.
\end{equation}
In the thermodynamical 
limit ($N,L\rightarrow \infty $ with $N/L=n$ ) ${\cal O}_{N,L}$ goes like 
 to $n (1-n)$, which is nonvanishing as $|j-l|\rightarrow
\infty .$ In other words the number state exhinits ODLRO, and thus is
superconducting.

The two-site reduced density matrix has the form
(\ref{eq:reduce}) in which
\begin{eqnarray}
u&=& \frac{(L-N)(L-N-1)}{L (L-1)},\nonumber \\
  z&=&{\cal O}_{N,L},\quad 
v=\frac{N (N-1)}{L(L-1)}.
\end{eqnarray}
From these relations and Eq. (\ref{eq:conccc})
one finds
\begin{equation}
C=2\left\{ {\cal O}_{N,L}-\left[ {\cal O}_{N,L}\frac{(N-1)(L-N-1)}{L(L-1)}%
\right] ^{1/2}\right\} .
\label{eq:C}
\end{equation}
Notice that the formula above could have been directly obtained from Ref (\cite
{WangKlaus}) where  the entanglement
between any pair of qubits in a Dicke state has been  computed.
Indeed by means of the  the mapping (\ref{mapp}), one can identify the number state with 
 the usual Dicke state

In the thermodynamical limit 
one has $u\mapsto (1-n)^2,\,v\mapsto n^2,\, z={\cal O}_{N,L}\mapsto n (1-n)$
thus, from Eq. (\ref{eq:conccc})  the entanglement  becomes zero.
So we see that the pairwise quantum entanglement does not exists although we
have $\eta $-pairing superconductivity in the number state. We 
finally notice that  in the  $\eta$-pair coherent states 
discussed in \cite{Solomon}, there is ODLRO, but being a products  there obviously
no entanglement.

\section{ Conclusions}

%%%%%%%%%%%%%%%%%%%%%%%%%%%%%%%%%%%%%%%%%%%%%%%%%%%%%%%%%%%%%%%%
The Fock space of many local fermionic modes can be mapped isomorphically
onto a many-qubit space. Using such a mapping we studied entanglement
between pairs of (spinless) fermionic modes. This has been done both for the
eigenstates and the thermal state for a model of free fermions hopping in a
lattice (entanglement between local modes as function of temperature and
filling). In the free fermionic model we analysed entanglement between local 
modes as function of temperature and filling. 
In particular we found that above a threshold 
temperature the thermal state becomes separable.

We studied the relations between pairwise entanglement and the superconducting
correlation in both the BCS-like model and the $\eta$-pair superconductivity. 
For the BCS-model, that a finite value of the superconducting order parameter is
required to obtain entanglement in the thermal state. Notice that this last
statement establishes a direct connection bewteen quantum entanglement and a
real phase transition \cite{Dorit}. For  $\eta$-pair superconductivity 
we found  that  pairwise entanglement is not a necessary condition for the
superconductivity. 

Despite their simplicity, our results seem to suggest that the quantum
information-theoretic relevant notion of quantum entanglement can provide
useful physical insights in the physics of many-body systems of
indistinguishable particles. 

\acknowledgments
We thank D Lidar for useful comments. This work has been supported by the
European Community through the grant IST-1999-10596 ( Q-ACTA).

\begin{table}[tbp]
\caption{The concurrence for three special values of mean fermion number.}
\begin{tabular}{cccc}
$\langle \hat{N}\rangle $ & $\mu $ & ${C}$ &  \\ \hline
$0$ & $-\infty$ & $0$ &  \\ 
$1$ & $0$ & $\max\left[\frac{\sinh(2 \beta |t|)-1}{\cosh(2\beta t)+1}%
,0\right]$ &  \\ 
$2$ & $\infty$ & $0$ & 
\end{tabular}
\end{table}

\begin{figure}[tbp]
\epsfig{width=10cm,file=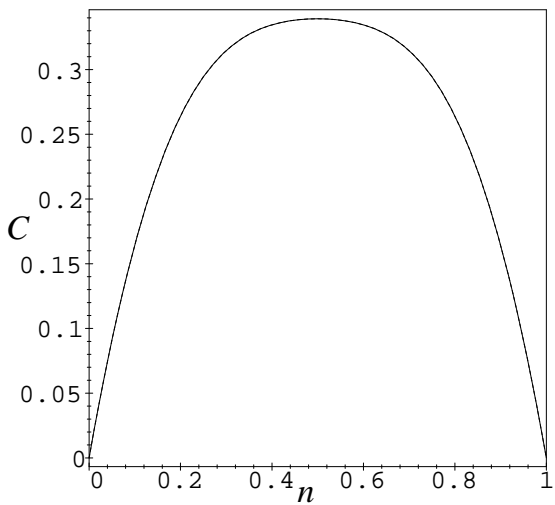}
\caption{The concurrence as a function of the filling $n$ in an infinite lattice}
\end{figure}

\newpage

\begin{figure}[tbp]
\epsfig{width=10cm,file=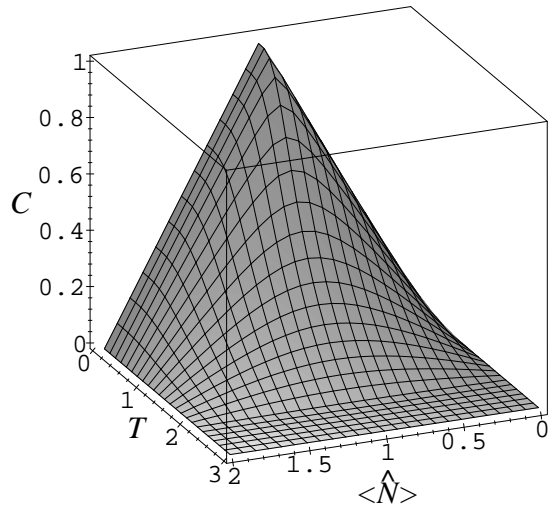}
\caption{The concurrence as a function of the temperature and the mean
fermion number for $L=2$ and $t=1$.}
\end{figure}

\begin{figure}[tbp]
\epsfig{width=10cm,file=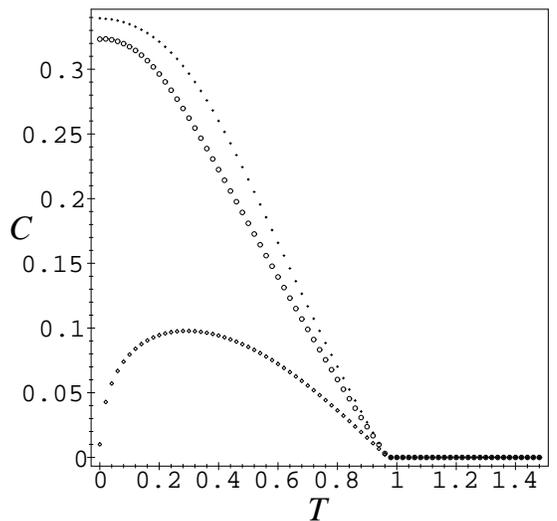}
\caption{The concurrence as a function of the temperature for different $\mu$%
: $\mu=0.1$ (cross points) , $\mu=1.0$ (circle points), and $\mu=2.0$
(diamond points). The parameter $L=100$.}
\end{figure}

\begin{figure}[tbp]
\epsfig{width=10cm,file=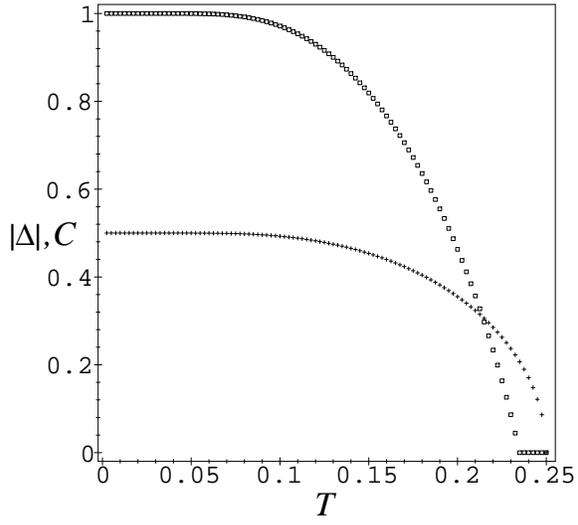}
\caption{The order parameter (cross points) and the concurrence (box points)
as a function of the temperature. The parameter $\epsilon_k=0$.}
\end{figure}

\begin{figure}[tbp]
\epsfig{width=10cm,file=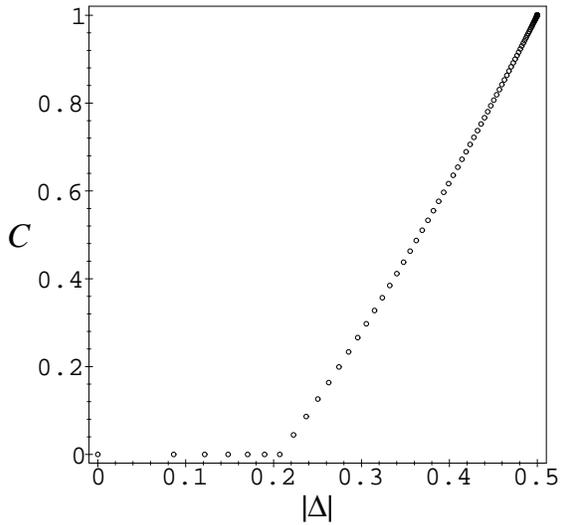}
\caption{The concurrence against the order parameter. The parameter $%
\epsilon_k=0$.}
\end{figure}


\begin{references}
\bibitem{hf}  This is immediately seen from the equation Tr$[\rho
_k\,(n_k+n_{-k})]=1-2\cot \theta _k\,|\Delta _k|.$
\end{references}
\end{document}